\documentclass[acmlarge]{acmart}

\AtBeginDocument{%
  }

\setcopyright{acmcopyright}
\copyrightyear{2023}
\acmYear{2023}
\acmDOI{}

\acmConference[Something]{1st International Workshop on Explainable AI for the Arts (XAIxArts), ACM Creativity and Cognition (C\&C) 2023}{}{}
\acmBooktitle{1st International Workshop on Explainable AI for the Arts (XAIxArts), ACM Creativity and Cognition (C\&C) 2023} 
\acmISBN{}

\begin{document}

\title[Framework for Matching Creative Tasks with Explanation Mediums]{Explaining the Arts: Toward a Framework for Matching Creative Tasks with Appropriate Explanation Mediums}

\author{Michael Clemens}
\email{michael.clemens@utah.edu}
\orcid{0000-0002-4507-8421}
\affiliation{%
  \institution{University of Utah}
  \streetaddress{50 S Central Campus Dr}
  \city{Salt Lake City}
  \state{UT}
  \country{USA}
  \postcode{46304}
}

\renewcommand{\shortauthors}{Clemens}

\begin{abstract}
  Although explainable computational creativity seeks to create and sustain computational models of creativity that foster a collaboratively creative process through explainability, there remains little to no work in supporting designers when exploring the explanation medium.  While explainable artificial intelligence methods tend to support textual, visual, and numerical explanations, within the arts, interaction mediums such as auditorial, tactile, and olfactoral may offer more salient communication within the creative process itself.  Through this research, I propose a framework to assist designers of explainable user interfaces in modeling the type of interaction they wish to create using explanations.
\end{abstract}

\received{2 May 2023}

\maketitle

\section{Introduction}
Neural networks and deep learning are widely used for co-creative applications, but their explainability is a major challenge \cite{Llano2022-bg}. Most explainable artificial intelligence (XAI) research focuses on explaining predictions and decisions in domains such as healthcare, finance, law, and criminal justice \cite{mittelstadt2019explaining}. My work, however, explores explainability within the artistic domains. XAI methods typically use visual, textual, or numerical explanations, but more than these may be needed for artistic contexts. Explainability in the arts is often related to intentionality and autonomy, which are considered essential for attributing creativity to an artificial agent \cite{collingwood1958principles,varshney2020limits}. Recent surveys further indicate that people expect artificial systems to exhibit novelty, quality, intentionality, and autonomy to be considered creative \cite{Das2022-xu}.

Explainable computational creativity is a branch of XAI that aims to build models that enable bi-directional communication between the user and the system. Based on HCI and creativity literature, Bryan-Kinns et al. \cite{bryan2022exploring} proposed a framework to analyze AI in the interactive arts along three dimensions: the role of AI, the interaction with AI, and the common ground with AI. They claim that the explainability of a creative AI depends on these aspects. For example, more explanation may be necessary when the process is more collaborative, requiring more engagement and grounding. Also, more contact with the agent helps users to learn about and infer the knowledge and understanding of the creative AI.  My research extends their work by contributing a framework that helps designers select the best explanation medium per the interaction type alongside the artistic domain.

Current efforts within the field of creative AI have created a surge of interest in these autonomous systems, however it's important to note there are various ways to describe the type of interaction, as previously mentioned, between the human and the computationally creative system.  These types of interactions include: autonomous systems, creativity support tools (CSTs), and co-creative systems.  Autonomous systems produce creative artifacts without any interaction from the user \cite{colton2015painting}. CSTs refer to tools and apps that are designed to aid in the user’s creativity \cite{compton2015casual}. Shneiderman \cite{shneiderman2007creativity} defines CSTs as tools that enhance the creative thought of users and enable them to be both productive and innovative.  Co-creation is the process of collaboratively creating meaningful objects or activities, enabled by users’ ability to share emotions, experiences, and ideas without needing specific guidance or predetermined strategies from a central authority \cite{giaccardi2008creativity}.  I will use Karimi’s et al. definition of co-creativity and explicitly define the concept of co-creation as the “interaction between at least one AI agent and at least one human where they take action based on the response of their partner and their conceptualization of creativity during the co-creative task” \cite{karimi2018evaluating}.

While evaluating the system of interaction supported by the creative agent, it is also imperative to assess for whom the explanation is provided \cite{ehsan2020human}.  The target audience determines the most effective method for describing the why behind the decisions.  Two dimensions are often mentioned within XAI research concerning user characteristics: AI literacy \cite{long2020ai} and mastery of the domain.  Although some work mentions the concept of experts in terms of AI experts, I introduce the concept of experts and novices to qualitatively assess a user's agency and experience with the respective artistic domain.  Ehsan et al. \cite{ehsan2021explainable} found that AI literacy significantly affects how numerical explanations were assessed, and each group (non-AI and AI experts) had unique perceptions of what determined a human-like explanation.  In a music production context, Clemens et al. \cite{clemens2022case} found that novice users were enthusiastic about using the creative agent in their workflow, while experts were concerned about the creative agent's impact on their creative autonomy.  The drastic differences between these user groups' views of explanations necessitate their inclusion while creating this framework.

The final dimension I wish to explore within the framework is the artistic outlet where creative interaction occurs. Within the artistic domain of music, there are several forms of artistic expression, including composing, performing, and producing. Although all of these fall under the domain of music, they are very different types of artistic expressions in how they are experienced. Interaction modalities such as CSTs and co-creativity may better support composing and performing, while performing is more apt for autonomous systems such as Shimon \cite{Hoffman2010-vx}.

\section{Research Objective}
\pagestyle{plain}

My research aims to create a framework that designers of explainable user interfaces can use to explore the most salient explanation medium for their targeted interaction paradigm with the creative system. Figure 1 presents an example of the framework for visually intensive artistic expressions. Although only one type of artistic expression is listed here, the completed framework would also include artistic expressions more focused on auditory, haptic, olfactory, and gustatory. Although particular art forms will include aspects of each of these, such as drawing includes both visual and haptic aspects, there is a primary focus on the visual portion rather than the haptic aspect of the art itself. The distinctions of these art forms will be based on the most intensive sense being used within the expression.

\begin{figure}[ht]
    \centering
    \includegraphics[width=1\textwidth]{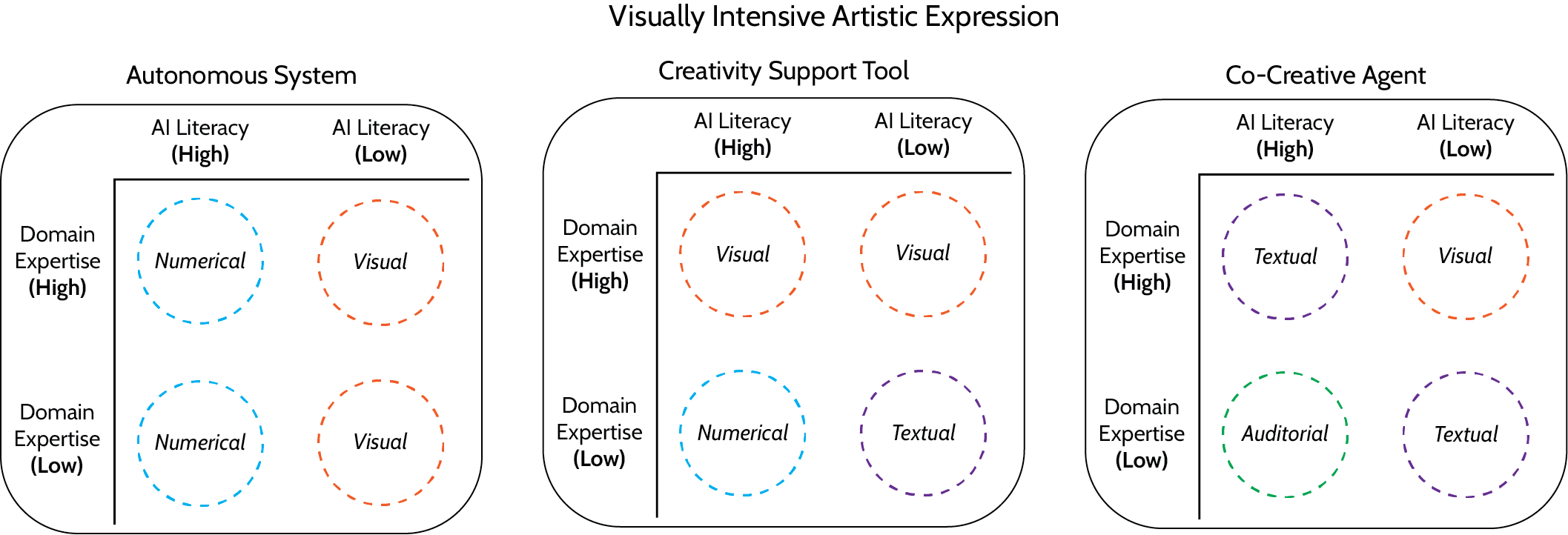}
    \caption{An example of the resulting explanations that would prove most suitable for designing explanations based on AI literacy and domain expertise of the user.  These medium choices are not grounded in data and are only there to demonstrate some possibilities.}
    \label{fig:research-questions}
\end{figure}

Figure 1 provides an example based on the two user dimensions: AI literacy and domain expertise, along with the appropriate mediums for each of the four created quadrants. Examples of the most salient medium are also displayed within each type of creative agent. Only three types of mediums are shown, but other types of medium could include auditorial, olfacorial, haptic, or gustoral.

The broader work in this area aims to develop a framework to measure an explanation's effectiveness in computational creative applications. One of the challenges is that there remains no universal standard for evaluating explanations in different creative domains or for differing user groups. This framework addresses this problem by explicitly defining the interaction between the user and the creative agent when explanations are provided. The next step in this research endeavor is to test this proposed framework on an artistic domain that uses senses not commonly used in XAI research, such as sound, and see if explanations in those modalities are more helpful for users than visual, textual, or numerical ones.

\bibliographystyle{xai-arts-reference-style}
\bibliography{main}

\end{document}